\documentclass[prl, nofootinbib, aps, twocolumn, 
showpacs, preprintnumbers, amsmath, amssymb, floatfix,superscriptaddress]{revtex4-1}
\usepackage{graphicx}
\usepackage{xspace}
\usepackage{booktabs}
\usepackage{bbm}
\usepackage{bm}
\usepackage[colorlinks=true,allcolors=blue]{hyperref}

\newcommand{\be}{\begin{equation}}
\newcommand{\ee}{\end{equation}}
\newcommand{\benn}{\begin{equation*}}
\newcommand{\eenn}{\end{equation*}}
\newcommand{\bse}{\begin{subequations}}
\newcommand{\ese}{\end{subequations}}

\renewcommand\({\left(}
\renewcommand\){\right)}

\newcommand{\exclude}[1]{}

\begin{document}
%
%
\preprint{NORDITA-2019-038}
\preprint{MIT-CTP-5116}
\preprint{MPP-2019-83}
%
%
%
\title{Tunable axion plasma haloscopes}
 
\author{Matthew Lawson}
\affiliation{The Oskar Klein Centre for Cosmoparticle Physics,
Department of Physics,
Stockholm University, AlbaNova, 10691 Stockholm, Sweden}\affiliation{Nordita, KTH Royal Institute of Technology and
Stockholm
  University, Roslagstullsbacken 23, 10691 Stockholm, Sweden}
\author{Alexander J.~Millar}
\affiliation{The Oskar Klein Centre for Cosmoparticle Physics,
Department of Physics,
Stockholm University, AlbaNova, 10691 Stockholm, Sweden}
\affiliation{Nordita, KTH Royal Institute of Technology and
Stockholm
  University, Roslagstullsbacken 23, 10691 Stockholm, Sweden}
\author{Matteo Pancaldi}
\affiliation{
Department of Physics,
Stockholm University, AlbaNova, 10691 Stockholm, Sweden}

\author{Edoardo Vitagliano}
\affiliation{Max-Planck-Institut f\"ur Physik (Werner-Heisenberg-Institut),
F\"ohringer Ring 6,\\ 80805 M\"unchen, Germany}
\author{Frank Wilczek}
\affiliation{The Oskar Klein Centre for Cosmoparticle Physics,
Department of Physics,
Stockholm University, AlbaNova, 10691 Stockholm, Sweden}
\affiliation{Nordita, KTH Royal Institute of Technology and
Stockholm
  University, Roslagstullsbacken 23, 10691 Stockholm, Sweden}
  \affiliation{Center for Theoretical Physics, Massachusetts Institute of Technology, Cambridge, Massachusetts 02139 USA}
  \affiliation{T. D. Lee Institute, Shanghai, China}
  \affiliation{Wilczek Quantum Center, Department of Physics and Astronomy,
Shanghai Jiao Tong University, Shanghai 200240, China}
\affiliation{Department of Physics and Origins Project, Arizona State University, Tempe AZ 25287 USA}

\begin{abstract}
	We propose a new strategy to search for dark matter axions using tunable cryogenic plasmas. Unlike current experiments, which repair the mismatch between axion and photon masses by breaking translational invariance (cavity and dielectric haloscopes), a plasma haloscope enables resonant conversion by matching the axion mass to a plasma frequency. A key advantage is that the plasma frequency is unrelated to the physical size of the device, allowing large conversion volumes. We identify wire metamaterials as a promising candidate plasma, wherein the plasma frequency can be tuned by varying the interwire spacing. For realistic experimental sizes we estimate competitive sensitivity for axion masses $35-400\,\mu$eV, at least.   
\end{abstract}

%

\maketitle
%

\section{Introduction}
One of the most pressing problems in cosmology is the composition of dark matter (DM). Recently the axion has been pushed towards the limelight as a DM candidate. Consequently, there is increasing urgency in developing strategies to search for this elusive particle. Originally the axion was introduced via the Peccei-Quinn (PQ) mechanism to solve the absence of significant CP violation in quantum chromodynamics (QCD), which is otherwise puzzling: the strong CP problem. The PQ mechanism essentially replaces the CP violating phase $\theta$ with a field which has a potential minimum at $\theta=0$ (the axion)~\cite{Peccei:1977hh,Weinberg:1977ma,Wilczek:1977pj}. Here the strong CP problem is solved dynamically, with the axion relaxing to the bottom of its potential over cosmological time scales.  


Residual oscillations of the axion field, while very small, remain until the present day and act as cold dark matter~\cite{Marsh:2015xka}. Depending on the exact cosmological history a wide range of axion masses $m_a$ can provide the correct abundance of dark matter, from $10^{-6}$--$10^3\,\mu$eV. Cosmology allows us to consider two broad scenarios. The axion itself is a pseudo-Goldstone boson generated by the breaking of the PQ symmetry; whether or not this symmetry is restored after inflation sets the initial conditions for the axion field. If the symmetry is restored after inflation, in each causally separated region of space the initial angle $\theta_i$ adopts a different value. As our observable universe would consist of many such patches, we live in an ``averaged" universe where the dark matter abundance is simply set by $m_a$. In addition, due to large inhomogeneities in the axion field topological defects such as strings and domain walls may form. These topological defects lead to difficulty in calculating the preferred axion mass~\cite{Kawasaki:2014sqa,Vaquero:2018tib}, though recent calculations suggest $m_a=25.2 \pm 11.0\,\mu$eV~\cite{Klaer:2017ond,Buschmann:2019icd}. Note however that large log factors are still poorly understood and may significantly change this prediction~\cite{Gorghetto:2018myk,Kawasaki:2018bzv}.

In contrast, if the PQ symmetry is not restored after inflation then the same $\theta_i$ exists throughout the observable universe. The exact value cannot be predicted, though often $\theta_i={\cal O}(1)$ is considered ``natural", giving $10^{-1}\,\mu{\rm eV}\lesssim m_a\lesssim 100\,\mu\rm eV$.

Searches based on cavity resonators in strong magnetic fields~\cite{Sikivie:1983ip} like ADMX~\cite{Rybka:2014xca}, HAYSTAC~\cite{Brubaker:2016ktl}, CULTASK~\cite{Woohyun:2016}, KLASH~\cite{Alesini:2017ifp} or ORGAN~\cite{Goryachev:2017wpw} are optimal for $m_a\lesssim 10\,\mu\rm eV$. Significantly lower values of $m_a$ can be explored by LC circuits~\cite{Sikivie:2013laa,Kahn:2016aff} or nuclear magnetic resonance techniques like CASPER \cite{Budker:2013hfa}. Recently, dielectric haloscopes~\cite{TheMADMAXWorkingGroup:2016hpc,Millar:2016cjp} were introduced to tackle the higher mass parameter space ($m_a\gtrsim 40\,\mu$eV). However, the technique is still under development. Similarly, the high mass performance of cavity haloscopes relies on unproven techniques such as combining large numbers of cavities~\cite{Goryachev:2017wpw,Jeong:2017hqs,Melcon:2018dba}. 

In this Letter we consider the coupling of the axion to bulk plasmons, rather than photons. The resonant mixing of axions with plasmons was first noted in~\cite{Mikheev:1998bg,Das:2004ee,Ganguly:2008kh}, but it was generally ignored until recently~\cite{Millar:2017eoc,Visinelli:2018zif,Tercas:2018gxv}. Concurrent with this work the idea of plasma-shining-through-wall has been introduced~\cite{Mendonca:2019eke}. Here, we consider the mixing of axions to tunable plasmas which can operate at cryogenic temperatures, thus leading to the first practical proposal to search for axions using axion-plasmon resonance. 

One key advantage to such a scheme, as we will demonstrate, is that the resonant frequency of the experiment is decoupled from its physical dimensions, side-stepping one of the main difficulties of building high frequency cavity haloscopes. Thus we expect a significant increase in the generated signal at high frequencies, even for less resonant setups.

\section{\boldmath Axion-Plasmon resonance in finite media} 
Despite axion-photon mixing being noted almost immediately after the introduction of the axion, the concomitant axion-plasmon mixing has been generally neglected. Throughout this Letter we will use natural units with the Lorentz-Heaviside convention. The axion's interaction with electrodynamics is described by a term 
\begin{equation}\label{lagrangian}
{\cal L} \supset  -\frac{g_{a\gamma}}{4}F_{\mu\nu}\widetilde F^{\mu\nu}a, 
\end{equation} 
with the coupling strength between axions and photons governed by the dimensionful constant $g_{a\gamma}$. The axion's mass and couplings are given by the decay constant $f_a$, with
$m_af_a\sim m_\pi f_\pi$ and $g_{a\gamma}=-\frac{\alpha}{2\pi f_a}\,C_{a\gamma}$, with $C_{a\gamma}$ an ${\cal O}(1)$ model dependent number and $\alpha$ the fine structure constant. 
We use $F_{\mu\nu}$ to denote the electromagnetic
field-strength tensor, with the dual tensor
\smash{$\widetilde F^{\mu\nu} = \frac{1}{2}\varepsilon^{\mu\nu\alpha\beta}F_{\alpha\beta}$}. Axion dark matter is extremely cold ($v\sim 10^{-3}$), giving a correspondingly large de Broglie wavelength $\lambda_{\rm dB}=2\pi/m_a v_a$. Thus the axion acts essentially as a spatially constant classical field oscillating with a frequency $\omega_a=m_a$, given by $a(t)=a_0e^{-i\omega_a t}$. In the presence of an external magnetic field ${\bf B}_{\rm e}$ the only modification to Maxwell's equations to lowest order in $g_{a\gamma}$ is in  Amp\`ere's law,\footnote{Due to the smallness of the coupling we will only consider linear order effects in $g_{a\gamma}$.}
\begin{equation}
	{\bm \nabla}{\bm \times} {\bf H} - \dot {\bf D}   = g_{a\gamma}{\bf B_{\rm e}}\dot a \,.
\end{equation}
 The primary effect of the axion is to act like an oscillating current, driving the system at $\omega_a$. Alternatively, one can think of ${\bf B}_{\rm e}$ inducing a mixing between the axion and photon. Due to conservation of momentum and energy, the massive axion cannot convert to a massless photon in an infinite space. In order to allow conversion between the two particles, one must either overcome the difference in dispersion relation or break translation invariance. Cavity and dielectric haloscopes do the latter by introducing structure on the scale of the Compton wavelength, allowing for the momentum mismatch to be satisfied. However, it is also possible to match dispersion relations by tuning the mass of the photon (plasma frequency) to that of the axion. In this case one does not need to break translation invariance, allowing for systems to be much larger than the Compton wavelength. 
 \subsection{Infinite plasma}
Previously, analysis of axion-plasmon mixing has been limited to the case of an infinite homogenous medium. In this case Maxwell's equations become particularly simple; if we consider an $E$-field in a linear medium with dielectric constant $\epsilon$, we see that
	\begin{equation}
		{\bf E}   = -\frac{g_{a\gamma}{\bf B}_{\rm e} a }{\epsilon}=-g_{a\gamma}{\bf B}_{\rm e}a \(1-\frac{\omega_p^2}{\omega_a^2-i\omega_a\Gamma}\)^{-1}\, ,
\end{equation} 
	where in the last equality we have introduced a Drude model for the dielectric constant. The plasma frequency is denoted by $\omega_p$, with $\Gamma$ being the small damping rate which sets the plasmon lifetime.
	In the limit ${\rm Re}(\epsilon)\to0$ a resonance occurs, corresponding to matching the axion frequency $\omega_a$ to the plasma frequency $\omega_p$. Importantly, the resonant frequency is a property of the medium, rather than being a function of the size of the system, as is the case for cavity haloscopes.
	
\subsection{Plasma cylinder}
As any experiment must be finite in extent we will now turn our attention to bounded plasmas. In particular, we take the plasma to be bounded inside a conductive cylinder. Similar cases are often considered under the label of ``plasma waveguides", which permits a similar analysis~\cite{bellan_2006}. 

The length of the cylinder is not actually relevant to the $E$ and $B$-field distributions, as any mismatch at the end caps is simply met by surface charges given by $({\bf D}_2-{\bf D}_1)\cdot{\bf n}_{12}=\sigma_s$. We choose cylindrical coordinates $(r,\phi,z)$ and take ${\bf B}_{\rm e}$ to be in the $z$ direction, i.e., ${\bf B}_{\rm e}=B_{\rm e}{\bf \hat z}$. As the axion is taken to be spatially constant, translational symmetry in the $z$ direction implies that all fields are similarly constant in the $z$ direction. In anticipation of a thin wire metamaterial, we will assume that the medium only has a non-unity dielectric constant $\epsilon_z$ in one direction, aligned with the cylinder. More generally, the strong applied magnetic field will ensure a highly anisotropic medium response. 

The cylindrical symmetry allows us to directly solve the axion-Maxwell equations at a radius $r$ inside a cylinder of total radius $R$, giving
\begin{subequations}
\begin{eqnarray}
	{\bf E}_z&=&-\frac{ag_{a\gamma}B_{\rm e}}{\epsilon_z}+\frac{ag_{a\gamma}B_{\rm e}}{\epsilon_z}\frac{J_0(\sqrt{\epsilon_z}r\omega)}{J_0(\sqrt{\epsilon_z}R\omega)}\,,\label{eq:E1}\\
	{\bf B}&=&-\frac{ag_{a\gamma}B_{\rm e}}{\sqrt {\epsilon_z}}\frac{J_1(\sqrt{\epsilon_z}r\omega)}{J_0(\sqrt{\epsilon_z}R\omega)}{\bf\hat \phi}\, .\label{eq:B1}
\end{eqnarray}
\end{subequations}
where $J_i$ is a Bessel function of the first kind. To unpack these expressions, note that we are interested in bulk plasmons, where ${\rm Re}(\epsilon_z)=0$. As $J_0(0)=1$, behaviour in the centre of the medium is dominated by $J_0(\sqrt{\epsilon_z}R\omega)$. When $R\gtrsim\pi/\sqrt{\epsilon_z}\omega$ the second term in equation~\eqref{eq:E1} and only term in~\eqref{eq:B1} can be neglected at the centre. In other words a sufficiently large medium has a bulk that behaves exactly the same as in the infinite medium case. This relationship gives us a minimum size for a given haloscope. Note that, unlike cavity or dielectric haloscopes, the enhancement of the $E$-field is not related to boundary conditions: the same would occur for an infinite medium, or one surrounded by vacuum.

\begin{figure}[t!]                        
\includegraphics[width=8cm]{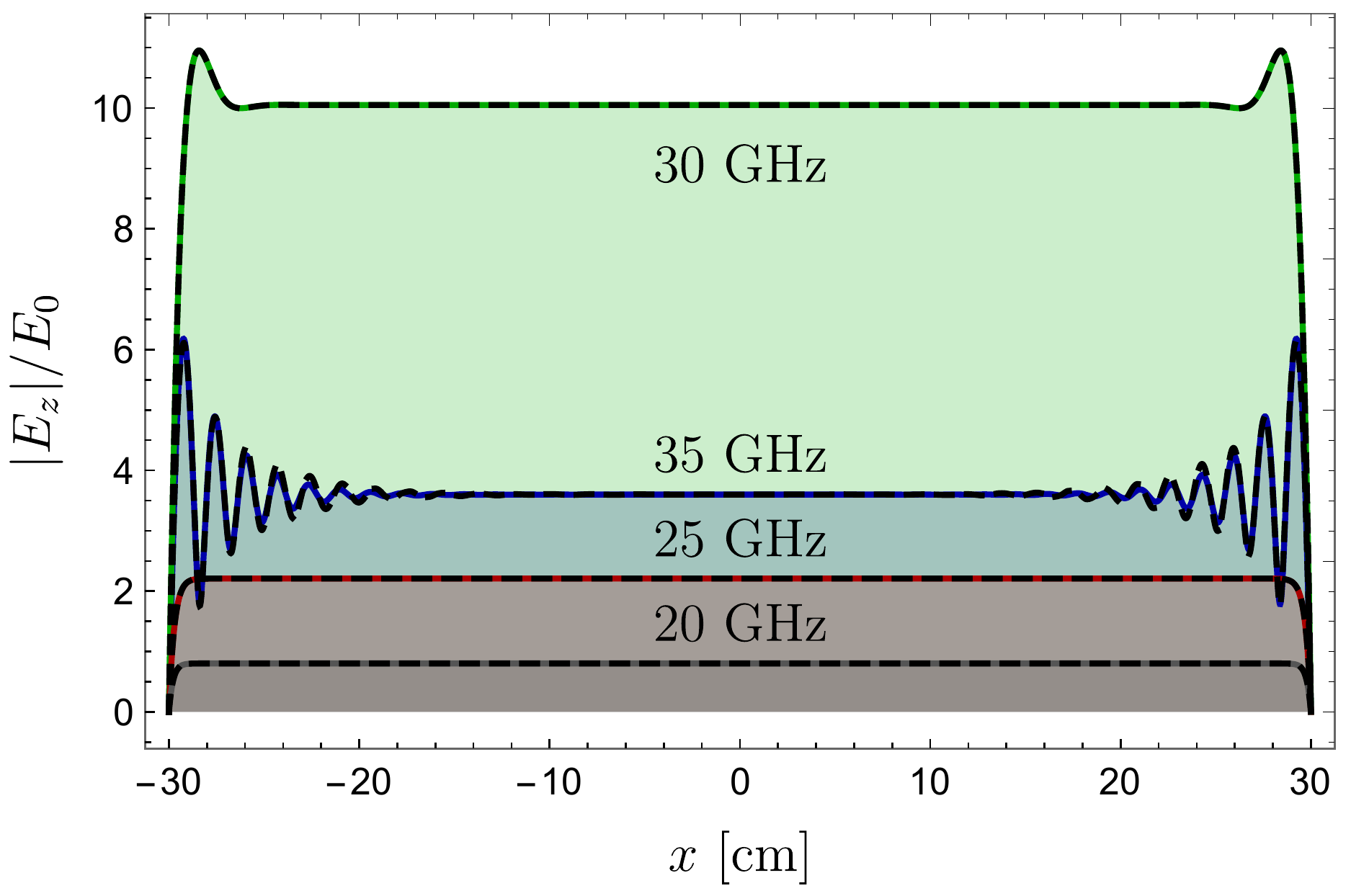}
\caption{Analytic and numerical calculation of $E$-field as a function of radius in an infinite, $30\,$cm radius cylinder of plasma contained inside conductive walls. We have defined $E_0=g_{a\gamma}{B}_{\rm e}a_0$. The plasma is characterised by $\omega_p=30\,$GHz and $\Gamma=10^{-1}\times\omega_p$, with axion frequencies $\omega=20,25,30~{\rm and}~35\,{\rm GHz}$ plotted. The coloured lines indicated the analytical calculation and the black dashed lines the numerical. Numerical calculations were performed in COMSOL for a $2.7\,$m long cylinder. }
\label{fig:plasma}
\end{figure}

To see these examples in action, in Fig.~\ref{fig:plasma} we plot the electric field profile in a $30\,$cm radius cylinder, with  $\omega_p=30\,$GHz and $\Gamma=10^{-1}\times \omega_p$. These are fairly conservative parameters, choosing a very lossy plasma. As expected, we see a large homogeneous region in the centre of the cylinder, with the edges having a small spike before rapidly vanishing at the edges.  We test our analytic calculations by comparison to numerical calculations performed in COMSOL~\cite{comsol}. The analytic and numerical calculations are essentially identical, showing a resonance when the axion and plasma frequencies match.  

\section{Wire metamaterials} 
In order to search for axions, we require materials that can be operated at cryogenic temperatures and have a tunable plasma frequency corresponding to the expected axion mass ($m_a\lesssim 60\,$meV)~\cite{Raffelt:2006cw,Chang:2018rso}. Several possibilities exist, including electron-poor semiconductors, field effect transistors, and Josephson junctions, which can often be tuned non-mechanically~\cite{PhysRevB.43.12630,cite-key,PhysRevB.83.014511}. As a prototypical example, we will instead focus on thin wire metamaterials~\cite{Pendry_1998}. One of the earliest proposed metamaterials, a volume filled with thin wires keeps many of the nice properties of metals (such as operating at cryogenic temperatures), however the plasma frequency of the system is vastly lower. We will consider homogeneously spaced wires aligned in the $z$ direction of radius $d$ and spacing $s$. It has been shown that as long as the wires are sufficiently thin ($\log(s/d)\gg 1$), they act as an effective medium with dielectric constant~\cite{Pendry_1998,PhysRevB.67.113103}
\begin{equation}
	\epsilon_z=1-\frac{\omega_p^2}{\omega^2-k_z^2+i\omega\Gamma}\,,
\end{equation}
where we have added in a small loss term.
As we are neglecting the axion velocity, $k_z=0$ so we can neglect spatial dispersion and simply recover a Drude-like model.
  
  To see that $\omega_p$ is much smaller than in a regular metal two effects must be considered. The first is the much lower average density of electrons $n_e$, and the second is the mutual inductance of the wires changing the effective electron mass $m_{eff}$. More explicitly these effects give~\cite{Pendry_1998}
\begin{equation}
	n_e=n\frac{\pi d^2}{s^2}\quad;\quad m_{eff}=\frac{e^2\pi d^2 n}{2\pi}\log\frac s d\,,
\end{equation}
where  $n$ is the density of electrons in the wires, and $e$ the electric charge. The corresponding plasma frequency is
\begin{equation}
	\omega_p^2=\frac {n_ee^2}{m_{eff}}=\frac{2\pi}{s^2\log(s/d)}\,\label{eq:plasmafreq}\,.
\end{equation}
We can easily see from \eqref{eq:plasmafreq} that the plasma frequency is essentially given by the spacing between the wires, meaning that for cm scale spacings $\omega_p=\cal{O}({\rm GHz})$. Further, as $\omega_p$ is a function of $s$, by changing the spacing of the wires it is possible to tune the plasma frequency. Thus we anticipate thin wire structures to be an ideal candidate plasma.

\section{Setup}
 A practical experimental realisation of this scheme must accommodate a few key features. First, the size and spacing of the wires must be such that the structure can be approximated as a medium, which condition obtains when $\log(s/d) \gg 1$ (that is, the diameter of the wires is much smaller than the spacing between the wires) \cite{Pendry_1998}.
 Copper wires with diameters as small as 10$\,\mu$m are commercially available, setting a soft limit of 3$\,$mm on the interwire spacing. This corresponds to a soft upper frequency limit of 100$\,$GHz. For higher frequencies the volume of the plasma will be limited also by $\lambda_{\rm dB}$. For haloscopes velocity effects start becoming noticeable for physical distances $\sim 20\%$ of $\lambda_{\rm dB}$~\cite{Irastorza:2012jq,Millar:2017eoc,Knirck:2018knd}.
 
 Second, the plasma must be placed in strong magnetic field, which limits the maximum practical diameter of the structure. Consequently, the maximum spacing between the wires is also limited. There are commercially available magnets with 7\,T fields and 60\,cm bore diameters, so taking this as a conservative upper limit, and requiring at least $\sim300$ evenly spaced wires, we find an interwire spacing of 3\,cm and a corresponding lower frequency limit of 8.8\,GHz.
 
 Third, the spacing between the wires must be tunable, while retaining a relatively high level of spacing homogeneity. However, tuning may be more easily realised than it seems at first glance. While, for simplicity, we have focused on isotropic wire arrays, this condition is not necessary; anisotropically spaced wires still give an isotropic effective medium~\cite{PhysRevB.67.113103}. As the plasma frequency comes from the mutual inductance between the wires the medium acts isotropically as long as the wires are arranged such that each wire feels the same mutual inductance as every other wire.
 
  Thus one very appealing geometry for tuning the plasma frequency is a series of planes of wires, with spacing adjusted in a single direction. We leave a full investigation of such mechanics to a technical design study. Assuming a tuning system which can cover more than a few percent of the total frequency range of the experiment  is realised, one could fabricate multiple inserts which could then be used for specific ranges of frequencies. Given that the overall size of the structure is largely independent of the operating frequency, such inserts could all be identical in external dimension, facilitating relatively fast swapping into and out of the magnet, electronic, and cryogenic systems. Depending on the details, such a tuning mechanism may further limit the high frequency range.
 
\begin{figure}[t!]                        
\includegraphics[width=.4\textwidth]{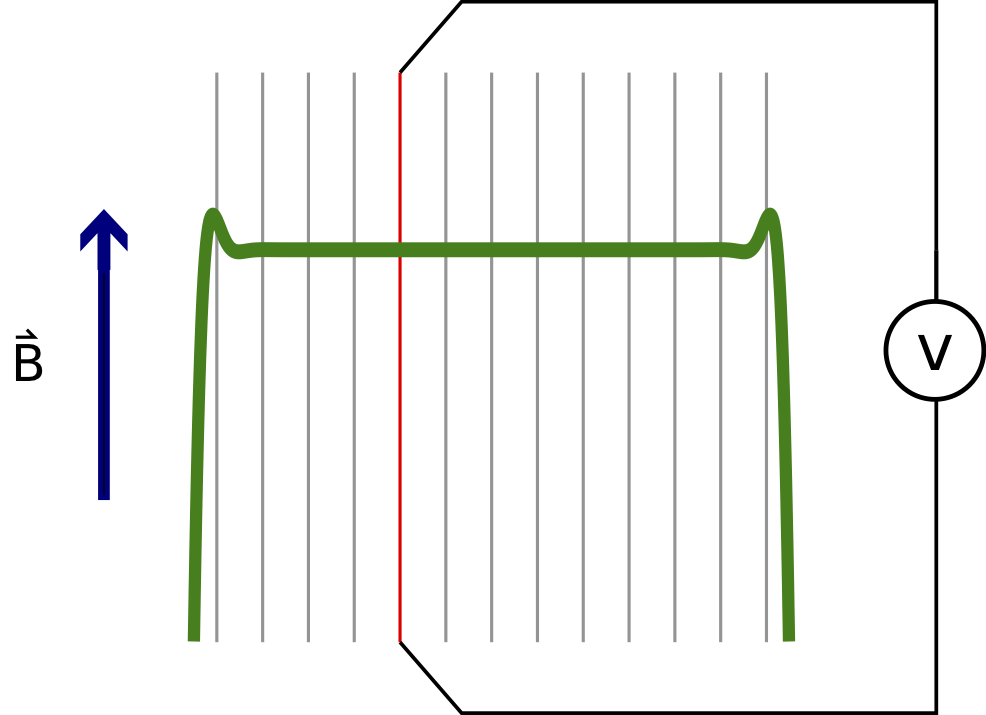} 
  \caption{A schematic representation of an experimental realization of the plasma haloscope. An array of wires with variable interwire spacing is placed in a strong uniform magnetic field. The array of wires (gray lines) acts as an effective medium with plasma frequency set (to leading order) by the wire spacing. The axion field excites a bulk plasmon in the wire metamaterial which can be detected by an antenna with the appropriate geometry (here represented by the red line). The green curve shows the absolute value of the electric field profile within the wire metamaterial. }
     \label{fig:diagram}
 \end{figure}

 Fig.~\ref{fig:diagram} shows a schematic representation of the essential features of the proposed experimental realization. An external magnetic field of order $10\,$T is applied to an array of parallel wires with tunable interwire spacing. In the presence of the magnetic field axions can induce bulk plasmons when the plasma frequency (set by the spacing) matches the axion frequency. As a result, there will be a
roughly uniform axial electric field oscillating sinusoidal in time. Antennas or pickup loops inside the structure
can be used to detect this oscillating electric field. An ultra-low-noise microwave amplifier then amplifies the voltage produced by the antenna. The wire assembly is electromagnetically shielded and cooled to cryogenic temperatures to reduce noise.

\section{Projected Reach}
To estimate the parameter space that could be explored by a plasma haloscope we must first calculate the power. In general, the power that can be extracted from the plasma is related to the loss rate of the plasma. For a Drude plasma, for small $\Gamma$ the loss rate is given by $P=\Gamma U$, where $U$ is the stored energy. This corresponds to a quality factor $Q=\omega/\Gamma$. Assuming a signal coupling efficiency factor $\kappa$, on resonance we can write the power in a form similar to that used for cavity haloscopes~\cite{Sikivie:1983ip},  
\begin{align}
	P&=\kappa {\cal G} V\frac{Q}{m_a}\rho_ag_{a\gamma}^2B_{\rm e}^2\,\label{eq:power},
\end{align}
where 
\begin{equation}
{\cal G}=\frac{\epsilon_z^2}{a_0^2g_{a\gamma}^2B_{\rm e}^2V}\frac{1}{2}\int \left (\frac{\partial(\epsilon_z\omega)}{\partial\omega}|{\bf E}|^2+|{\bf B}|^2 \right )dV\, .
\end{equation}
The local axion dark matter density is denoted $\rho_a$, with $0.45\,{\rm GeV}/{\rm cm}^3$ being the typical value used.
In $\cal G$ we have defined a kind of ``geometry factor", however unlike a traditional cavity haloscope ${\cal G}\to 1$ as $R\to \infty$. Further, as $g_{a\gamma}\propto m_a$ for the QCD axion the power actually increases at higher masses, given constant $Q$. 
Thus the primary advantage over traditional cavity haloscopes comes from the dramatically increased measurement volume, meaning that much more energy is stored in the system.  Because of this feature plasma haloscopes will be most useful in the high frequency regime. 

To calculate a scan rate, the signal to noise ratio (S/N) is given by Dicke's radiometer equation,
$S/N= (P/T_{\rm sys})\sqrt{\Delta t/\Delta \nu_a}$, where the system noise temperature is $T_{\rm sys}$ and the axion signal line width $\Delta \nu_a\sim 10^{-6}\nu_a$. The measurement time is denoted $\Delta t$, covering a frequency range $\sim \omega/Q$. Assuming that the tuning mechanism allows for rapid tuning and so can be neglected, it is simple to estimate the projected reach of a given experiment. Parametrizing the measurement time as $\Delta t=A\nu^p$, where $A$ and $p$ are independent of $\nu$ and given by equation~\eqref{eq:power}, we can integrate Dicke's formula between some frequencies $\nu_1,\nu_2$ to get the total scanning time
\begin{equation}
	t_{\rm scan}=\frac{QA}{p}\left (\nu_2^p-\nu_1^p\right)\,.
\end{equation}

\begin{figure}[t!]                         
\includegraphics[width=8cm]{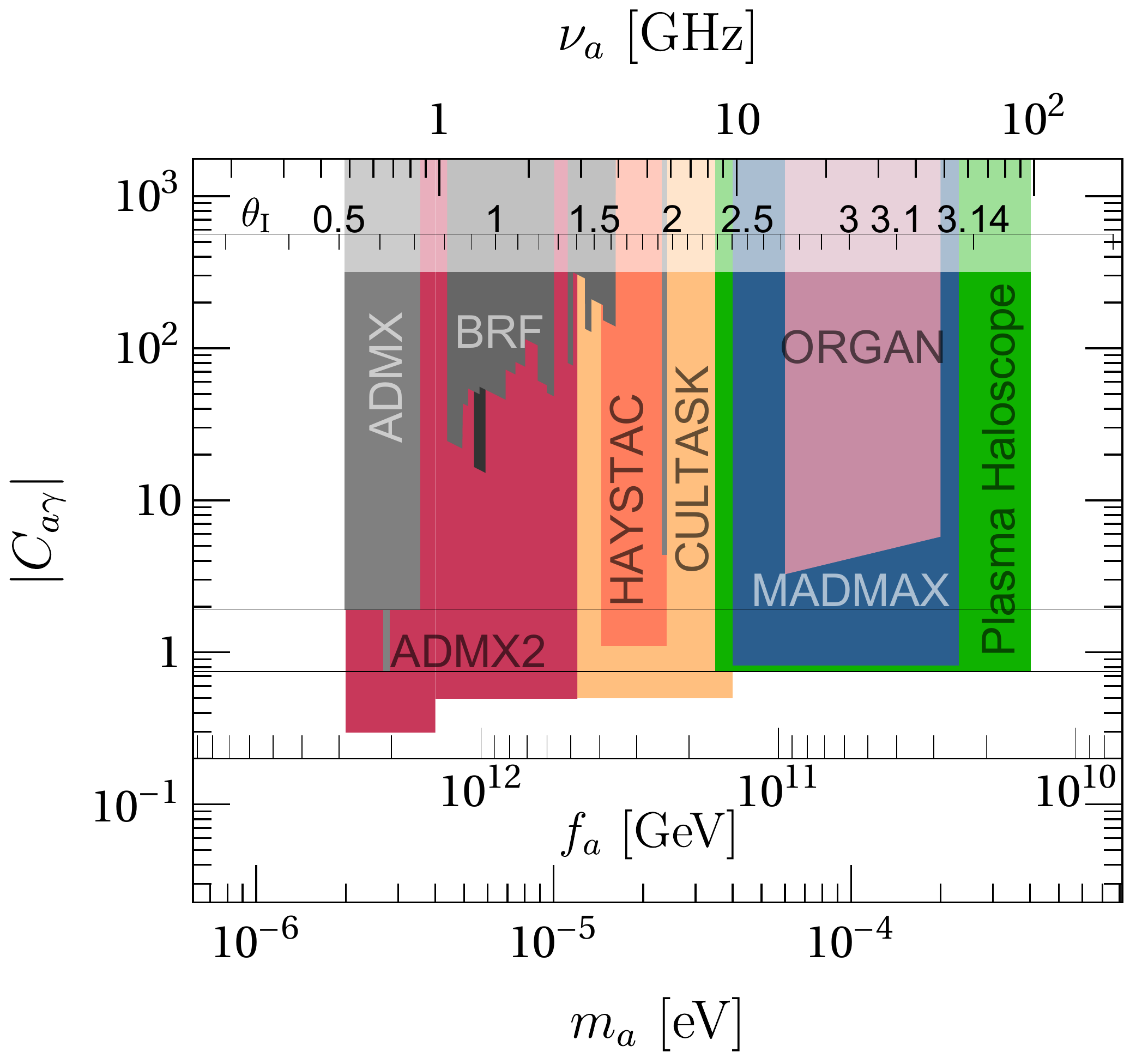} 
  \caption{In green we show the projected reach in the $\{m_a,|C_{a\gamma}|\}$ plane of a plasma with $Q=10^2$ and $V=0.8\, {\rm m}^3$ inside $10\,$T magnetic field. We assume a critically coupled, quantum limited detector with a 6 year measurement time. The black lines bracket the traditional axion model band, $0.746<|C_{a\gamma}|<1.92$. Existing limits are displayed in gray, with coloured regions corresponding to proposed experiments (red tones for cavities, blue for dielectric haloscopes)~\cite{Wuensch:1989sa,Hagmann:1990tj,Asztalos:2009yp,vanBibber2015Zaragoza,Brubaker:2016ktl,Woohyun:2016,Carosi2016Jeju,McAllister:2017lkb,Du:2018uak,TheMADMAXWorkingGroup:2016hpc,Zhong:2018rsr}. }
     \label{fig:reach} 
 \end{figure}

To put in some numbers, we assume a plasma with a conservative $Q=10^2$ and $V=0.8\, {\rm m}^3$ inside $10\,$T magnetic field. With such a conservative $Q$ and concomitant longer measurement time the assumption that the tuning time can be neglected should hold true. We require $S/N\geq 3$ over the full width half maximum ($\omega/Q$). To account for the decrease in $\lambda_{\rm dB}$ at high frequencies, we assume an 60\,cm bore magnet and limit the length to 30\% of $\lambda_{\rm dB}$.
 As shown in Fig.~\ref{fig:reach}, with quantum limited detection ($T_{\rm sys}=m_a$) the axion parameter space $35-400\,\mu{\rm eV}$ could be explored down to DFSZ~\cite{Dine:1981rt,Zhitnitsky:1980tq} ($|C_{a\gamma}|=0.746$) in 6 years. We have assumed that the readout system is critically coupled, i.e., $\kappa=0.5$. 

 Such a detection scheme is more easily reached at low masses, where Josephson parametric amplifiers operating with a dilution refrigerator have been shown to achieve near quantum limited detection~\cite{Zhong:2018rsr}. With a more modest detection system, such as a commercially available high electron mobility transistor operating in liquid helium ($T_{\rm sys}\sim 5\,$K) the higher mass range $100-160\,\mu{\rm eV}$ could be explored in a similar time frame (4 years). Such a scan would be complementary to the initial stages of MADMAX~\cite{TheMADMAXWorkingGroup:2016hpc}. Above $40\,$GHz new detection technology will be required regardless of the specific axion detection mechanism~\cite{Brun:2019lyf}.

Due to similar design limitations, thin wire metamaterials seem most optimal in a similar parameter space as dielectric haloscopes. Which technique will be more practical at a given frequency depends strongly on the detailed design of an experiment. A particular advantage of plasma haloscopes is that a solenoidal, rather than dipole, magnet can be used, simplifying the design and cost. In addition, plasma haloscopes may be more easily extended into even higher mass parameter space, as the technique will work as long as an appropriate medium is found. As mentioned above, several candidate materials exist with the possibility of high $Q$ and non-mechanical tuning and warrant further exploration.

\section{Conclusion}
In this Letter we have outlined a new axion detection technique using tunable cryogenic plasmas. By matching the axion and plasmon masses we induce resonant conversion between the two. Unlike existing haloscope designs axion-photon conversion is not caused by matching boundary conditions. Thus we open the possibility for dramatically increased measurement volumes, allowing significantly more power than a traditional cavity haloscope, especially at high frequencies. Using thin wire metameterials as a candidate plasma, we showed that plasma haloscopes provide a plausible alternative to dielectric haloscopes in the mass region $35-400\,\mu{\rm eV}$, at least.  Other candidate plasmas exist, which may allow the technique to be expanded into larger parameter spaces.

\begin{acknowledgements}
\section*{Acknowledgments}
The authors would like to thank Javier Redondo, Nathan Newman, Stefano Bonetti and J\'on Gudmundsson for helpful discussions. AM and ML are supported by the
European Research Council under Grant No.\ 742104. FW's work is supported by the U.S. Department of Energy under grant Contract  Number DE-SC0012567, by the European 
Research Council under grant 742104, and by the Swedish Research Council under Contract No. 335-2014-7424. EV acknowledges partial support by the Deutsche Forschungsgemeinschaft through Grant No. SFB 1258 (Collaborative Research Center “Neutrinos, Dark Matter, Messengers”) as well as by the European Union through Grant No. H2020-MSCA-ITN- 2015/674896 (Innovative Training Network “Elusives”).

\end{acknowledgements}

\providecommand{\href}[2]{#2}\begingroup\raggedright\endgroup

\end{document}